\documentclass[a4paper]{article}

\usepackage[utf8]{inputenc}

\usepackage{hyperref}
\usepackage[pdftex]{graphicx}
\usepackage{tikz}

\usepackage{xcolor}
\definecolor{colorone}{HTML}{116699}
\definecolor{colortwo}{HTML}{cccccc}
\definecolor{colorthree}{HTML}{cc6633}
\definecolor{textcolor}{HTML}{000000} 
\definecolor{airforceblue}{rgb}{0.36, 0.54, 0.66}
\definecolor{ao(english)}{rgb}{0.0, 0.5, 0.0}
\definecolor{britishracinggreen}{rgb}{0.0, 0.26, 0.15}
\definecolor{coolblack}{rgb}{0.0, 0.18, 0.39}

\usepackage[left=2.5cm,right=2.5cm,top=2.5cm,bottom=3cm]{geometry}

\usepackage[colorinlistoftodos]{todonotes}

\makeatletter
\def\@maketitle{%
  \newpage
  \null
  \vskip 2em%
  \begin{center}%
  \let \footnote \thanks
    {\Large\bfseries \@title \par}%
    \vskip 1.5em%
    {\normalsize
      \lineskip .5em%
      \begin{tabular}[t]{c}%
        \@author
      \end{tabular}\par}%
    \vskip 1em%
    {\normalsize \@date}%
  \end{center}%
  \par
  \vskip 1.5em}
\makeatother
\usepackage[affil-it]{authblk}

\usepackage{cleveref}

\begin{document}
\title{Search for high-energy neutrino emission from\\
Mrk 421 and Mrk 501\\
with the ANTARES neutrino telescope}

\author{M. Organokov\thanks{email: \texttt{mukharbek.organokov@iphc.cnrs.fr}; Corresponding author} \, and T. Pradier \\
on behalf of the ANTARES Collaboration}
\affil{Universit\'e de Strasbourg, CNRS, IPHC UMR 7178, F-67000 Strasbourg, France}


\date{Dated: \today}

\maketitle

%
\begin{minipage}{0.705\paperwidth}
\centering
\abstract{
ANTARES is the largest high-energy neutrino telescope in the Northern Hemisphere. This contribution presents the results of a search, based on the ANTARES data collected over 17 months between November 2014 and April 2016, for high energy neutrino emission in coincidence with TeV $\gamma$-ray flares from Markarian 421 and Markarian 501, two bright BL Lac extragalactic sources highly variable in flux, detected by the HAWC observatory. The analysis is based on an unbinned likelihood-ratio maximization method. The $\gamma$-ray lightcurves (LC) for each source were used to search for temporally correlated neutrinos, that would be produced in pp or p-$\gamma$ interactions. The impact of different flare selection criteria on the discovery neutrino flux is discussed. Plausible neutrino spectra derived from the observed $\gamma$-ray spectra in addition to generic spectra $E^{-2}$ and $E^{-2.5}$ are tested. 
}
%
\end{minipage}

\section{Introduction}
The very high energy (VHE; 0.1-100 TeV) extragalactic sky is dominated by emission from blazars~\cite{CerrHESS}, a class of a radio-loud AGN~\cite{AGNBeckmannPlot}. Two blazars, Mrk 421 and Mrk 501, the brightest and closest BL Lac objects known, at luminosity distances $\mathrm{d_L}$ = 134 Mpc with redshift z = 0.031 and $\mathrm{d_L}$ = 143 Mpc with redshift z=0.033 respectively. These blazars are the first and the second extragalactic objects discovered in the TeV energy band. This analysis focuses on the search of spatial/temporal correlation between neutrinos ($\nu$) detected by ANTARES and $\gamma$-ray emission from flares detected by HAWC from these blazars in the period Nov. 2014 - Apr. 2016, and reported in~\cite{HAWCdaily}. As the nearest blazars to Earth, both are excellent sources to test the blazar-neutrino connection scenario, especially during flares where time-dependent neutrino searches may have a higher detection probability.

\section{ANTARES and HAWC}
The ANTARES (Astronomy with a Neutrino Telescope and Abyss environmental RESearch) neutrino telescope~\cite{ANTARESfirstundersea} is a Cherenkov detector designed to search for high-energy neutrinos from astrophysical sources by detecting the Cherenkov light emission of neutrino-induced charged particles in the very deep waters of the Mediterranean Sea and most sensitive for neutrino energies 100 GeV $<E_{\nu}<$ 100 TeV. The ANTARES is located in the Mediterranean sea, 40 km off the coast of Toulon, France ($42^\circ 48'$ N $6^\circ 10'$), at a depth of 2,475 meters.
The HAWC (High-Altitude Water Cherenkov Observatory) gamma-ray observatory~\cite{HAWCdaily} is a Cherenkov detector designed to search for high-energy $\gamma$-rays  (100 GeV $<E_{\gamma}<$ 100 TeV) from astrophysical sources by detecting the Cherenkov light emission from charged particles in $\gamma$-ray induced air showers. HAWC is located at an elevation of 4,100 m above sea level on the flanks of the Sierra Negra volcano in the state of Puebla, Mexico ($18^\circ 99'$ N, $97^\circ 18'$ W).

\subsection{The ANTARES data set}
The ANTARES data set covers the same period of observation as HAWC: from November $26^{\mathrm{th}}$, 2014 until April $20^{\mathrm{th}}$, 2016 (MJD: 56988-57497) leading to an effective detector livetime of 503.7 days. The search relies on track-like event signatures, so only CC interactions of muon neutrinos are considered. The muon track reconstruction returns two quality parameters, namely the track-fit quality parameter, $\Lambda$, and the estimated angular uncertainty on the fitted muon track direction, $\beta$. Cuts on these parameters are used to improve the signal-to-noise ratio. To avoid biasing the analysis, it has been performed according to a blinding policy with the data have been blinded by time-scrambling. The final sensitivities are derived from the blinded dataset.

\subsection{The HAWC light curves of blazars}
\label{LCmrk}
HAWC has made clear detections of Mrk 421 and Mrk 501. In this analysis HAWC-300 data of first long-term TeV light curve studies with single-transit intervals are used~\cite{HAWCdaily}. The HAWC flare states for Mrk 421 and Mrk 501 with one day binning applied are shown in Fig.~\ref{fig:mrk421threshold} and Fig.~\ref{fig:mrk501threshold} respectively. These flare states are the Bayesian blocks derived from the LCs~\cite{HAWCdaily}. Several flare selection conditions considered: all flare states are taken as they are (long case);only those flare states are taken which pass the defined thresholds (short case): \textit{average flux}, \textit{average flux}+$\mathit{1\sigma}$, \textit{average flux}+$\mathit{2\sigma}$. 
\section{Time-dependent search method}
A search for neutrino candidades in coincidence with $\gamma$-rays from astrophysical sources is performed using an unbinned likelihood-ratio maximization method~\cite{blazarsfive,xrb,icrc17org}. 
The goal is to determine the relative contribution of background and signal components for a given direction in the sky and at a given time.
\begin{equation}
\centering
\mathrm{ln(\mathsf{L})}=\bigg(\sum_{\mathrm{i=1}}^{\mathrm{N}} \mathrm{ln[N_S S_i + N_B B_i]}\bigg) - \mathrm{[N_S + N_B]}
\end{equation} 

To perform the analysis, the ANTARES data sample is parametrized as two-component mixture of signal and background. The signal is expected to be small so that the full data direction can be used as an estimation of the background. $\mathrm{S_i}$ and $\mathrm{B_i}$ are defined as the probability density functions (PDF) respectively for signal and background for an event i, at time $\mathrm{t_i}$, energy $\mathrm{E_i}$, declination $\delta_\mathrm{i}$. As a result, $\mathrm{S_i}$ = $\mathrm{P_s}(\alpha_i) \cdot \mathrm{P_s(E_i)} \cdot \mathrm{P_s(t_i)}$ and $\mathrm{B_i}$ = $\mathrm{P_b(sin(\delta_i))} \cdot \mathrm{P_b(E_i)} \cdot \mathrm{P_b(t_i)}$. The parameter $\alpha_i$ represents the angular distance between the direction of the event i and direction to the source.
Additionally, $N_{S}$ and $N_{B}$ are unknown signal events and known background rate (a priori when building the $\mathsf{L}$) respectively. Since the signal is expected to be small, the total number of events N in the considered data sample can be treated as background.

The energy PDF for the signal events is produced according to the studied energy spectra: $E^{-2.0}$, $E^{-2.5}$, $E^{-1.0} \, \mathrm{exp}\left(-E \slash 1 \, \mathrm{PeV}\right)$ for both sources and extra $E^{-2.25}$ for Mrk 501 (as from the fit of the spectral shape of this source performed for same data in~\cite{HAWCspecEBL}). Worth noting that Mrk 421 is well described by the $E^{-2.0}$ spectrum.

\begin{figure}[!htb]
\centering
\includegraphics[width=0.825\textwidth]{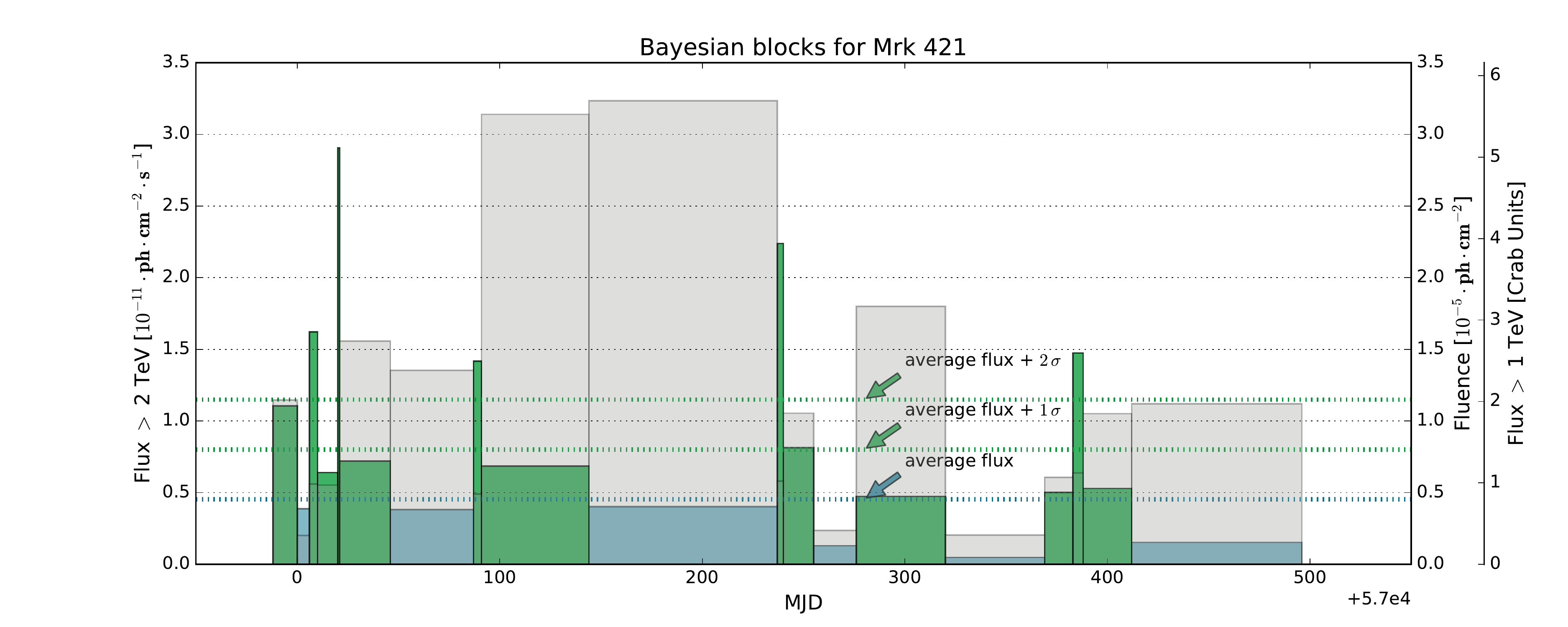}\\
\includegraphics[width=0.825\textwidth]{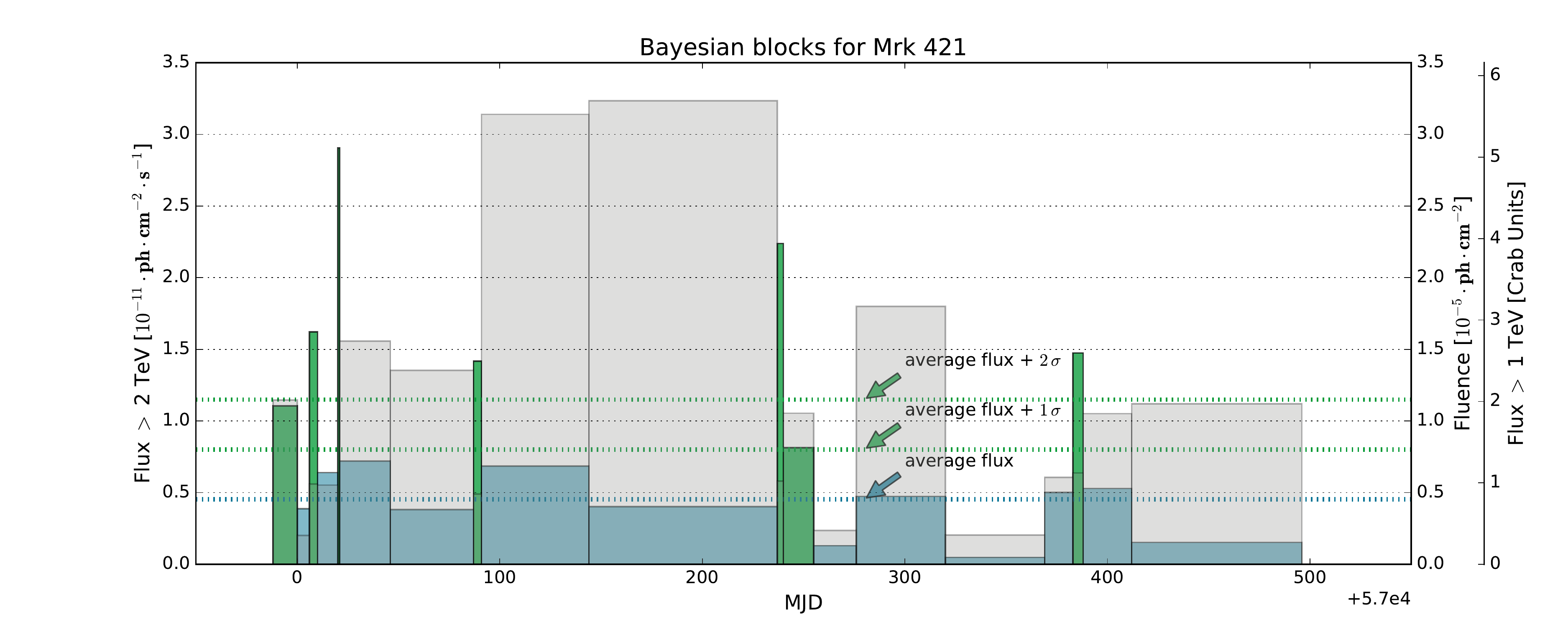}\\
\includegraphics[width=0.825\textwidth]{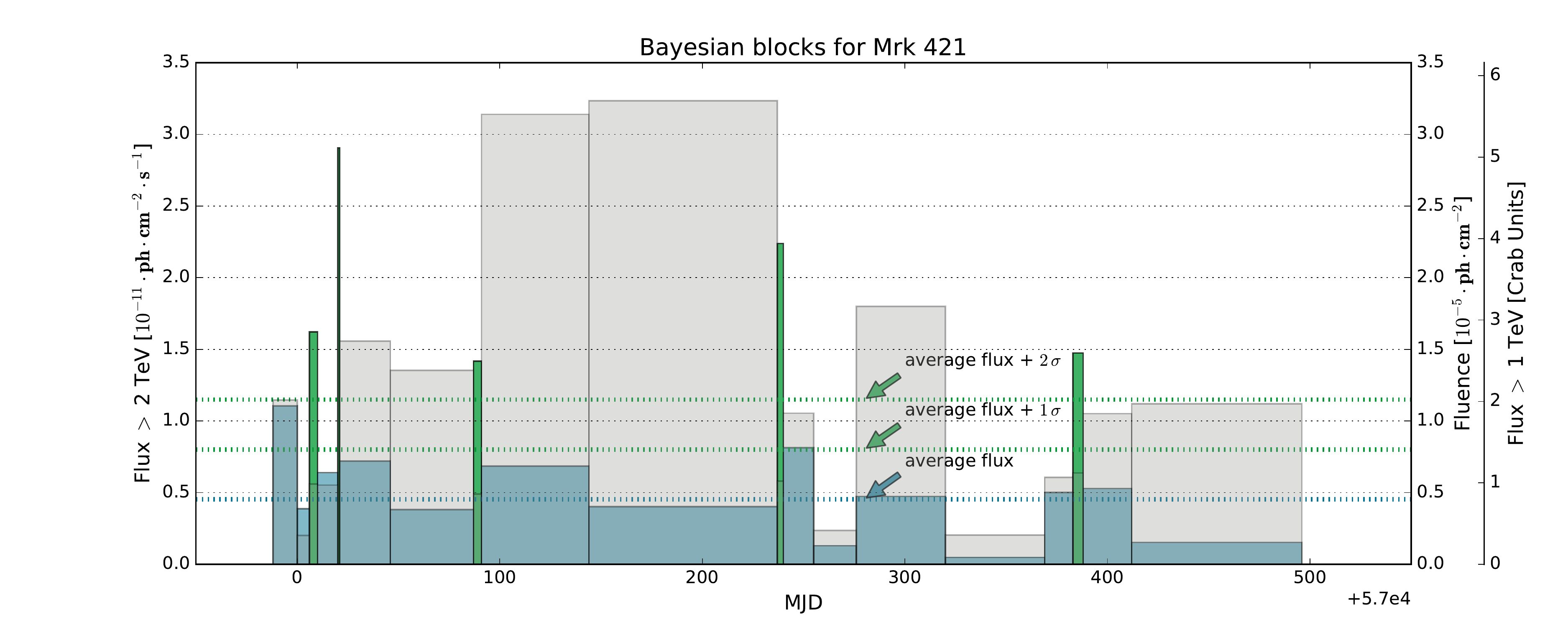}\\
\includegraphics[width=0.825\textwidth]{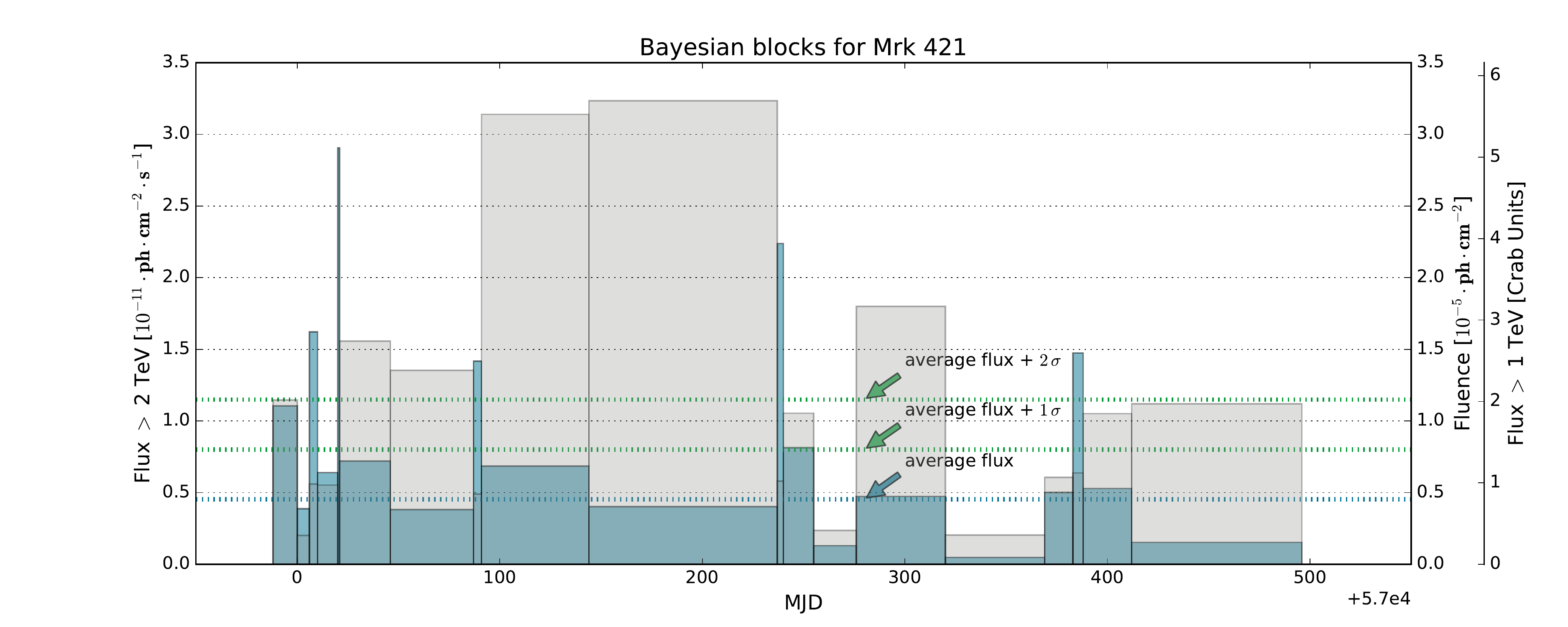}
\caption{Flare states for Mrk 421 vs threshold. The blue dotted line represent the average flux $\sim0.8$ CU (Crab Units); the green dotted lines represent the peak selection thresholds: \textit{average flux}, \textit{average flux +} $\mathit{1\sigma}$, \textit{average flux +} $\mathit{2\sigma}$. The bottommost plot shows the long case, the three upper plots show short case for \textit{average flux}, \textit{average flux +} $\mathit{1\sigma}$, \textit{average flux +} $\mathit{2\sigma}$ respectively. The left axes represent the units of the fluxes, the right-right axis represent the fluxes in corresponding CU. The right-left axes represent the units of fluences shown as shaded grey areas.}
\label{fig:mrk421threshold}
\end{figure}

\begin{figure}[!htb]
\centering
\includegraphics[width=0.825\textwidth]{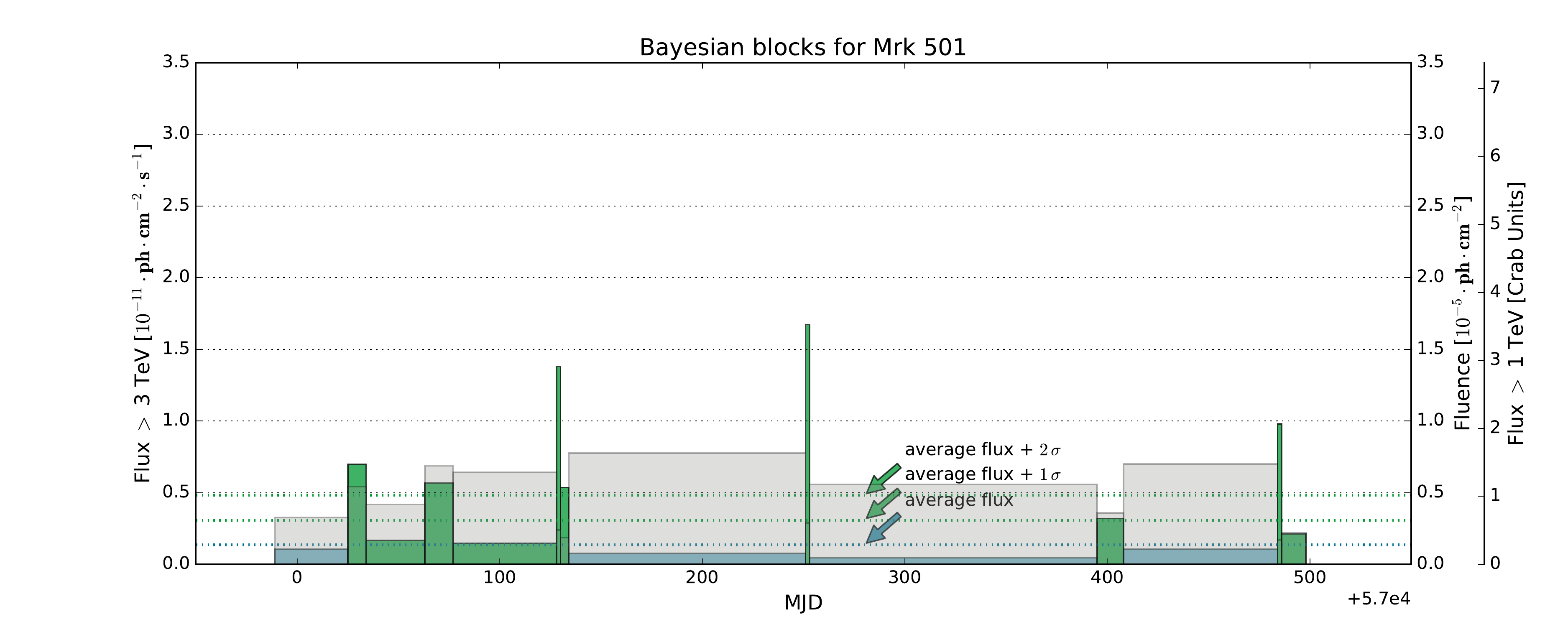}\\
\includegraphics[width=0.825\textwidth]{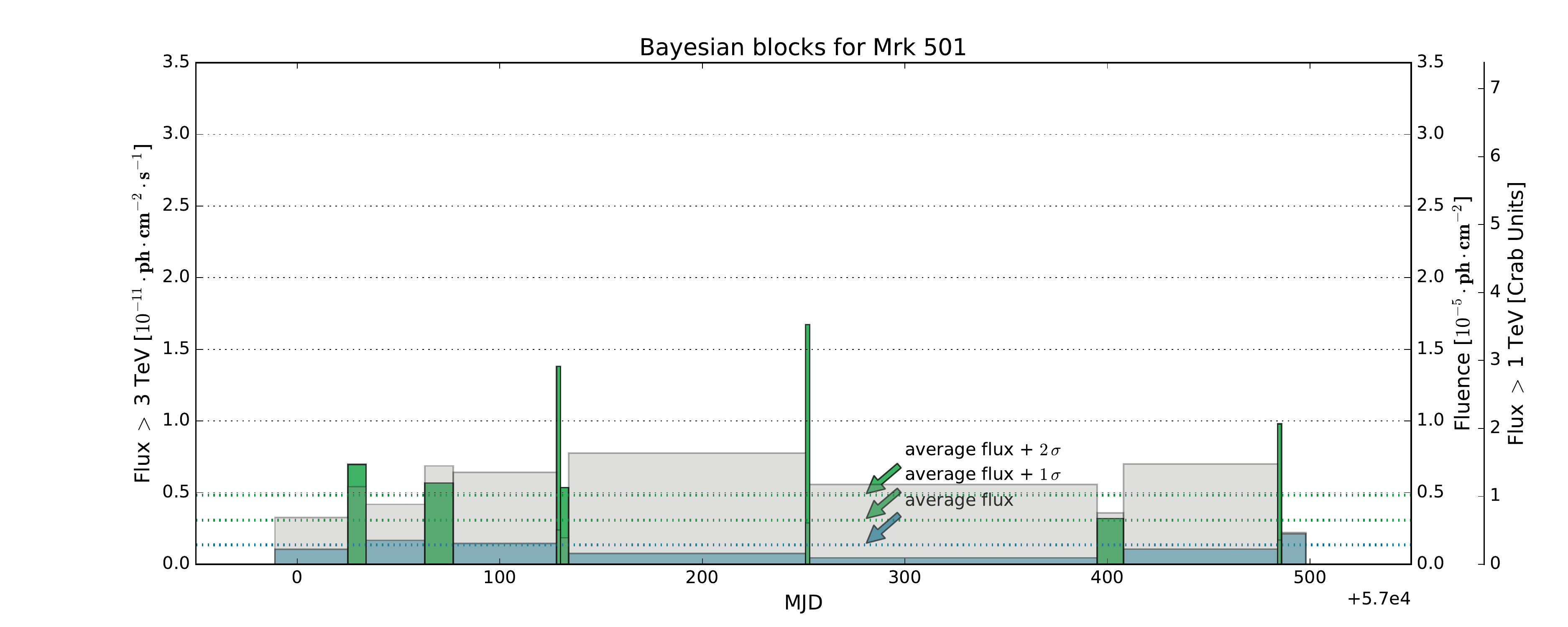}\\
\includegraphics[width=0.825\textwidth]{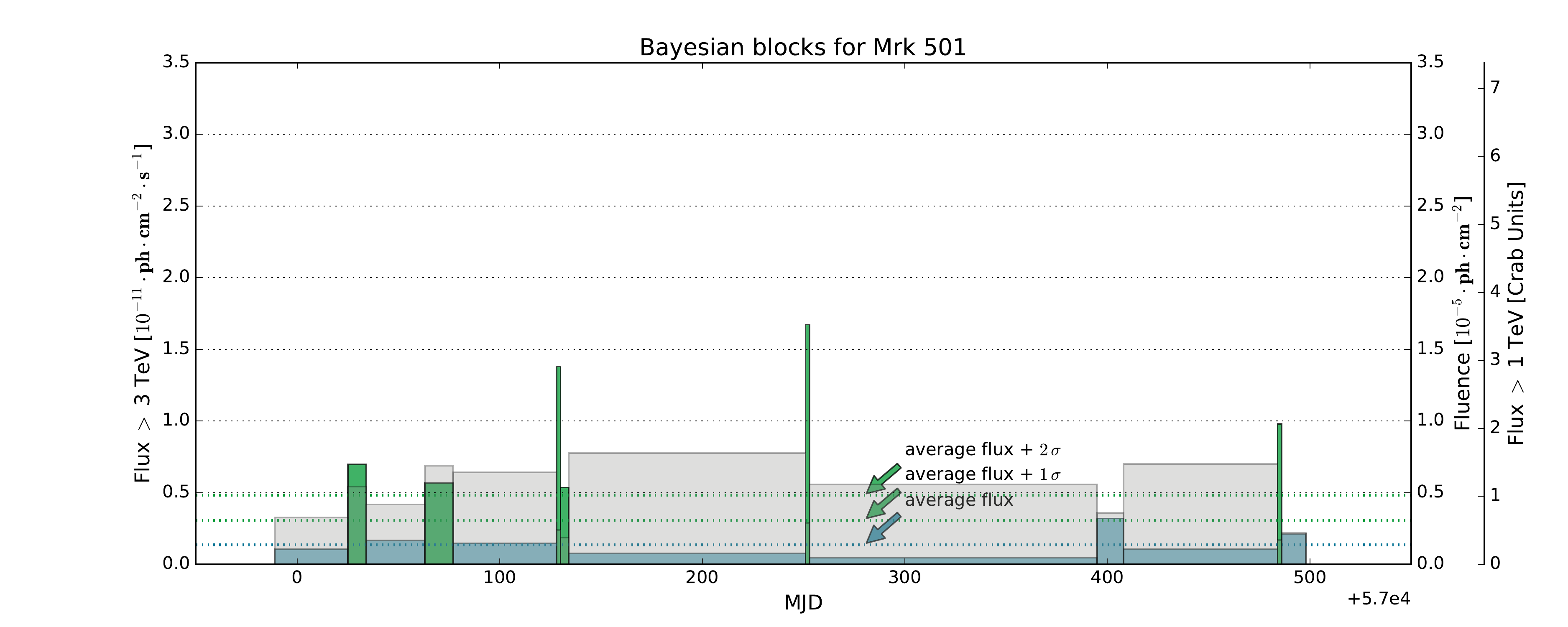}\\
\includegraphics[width=0.825\textwidth]{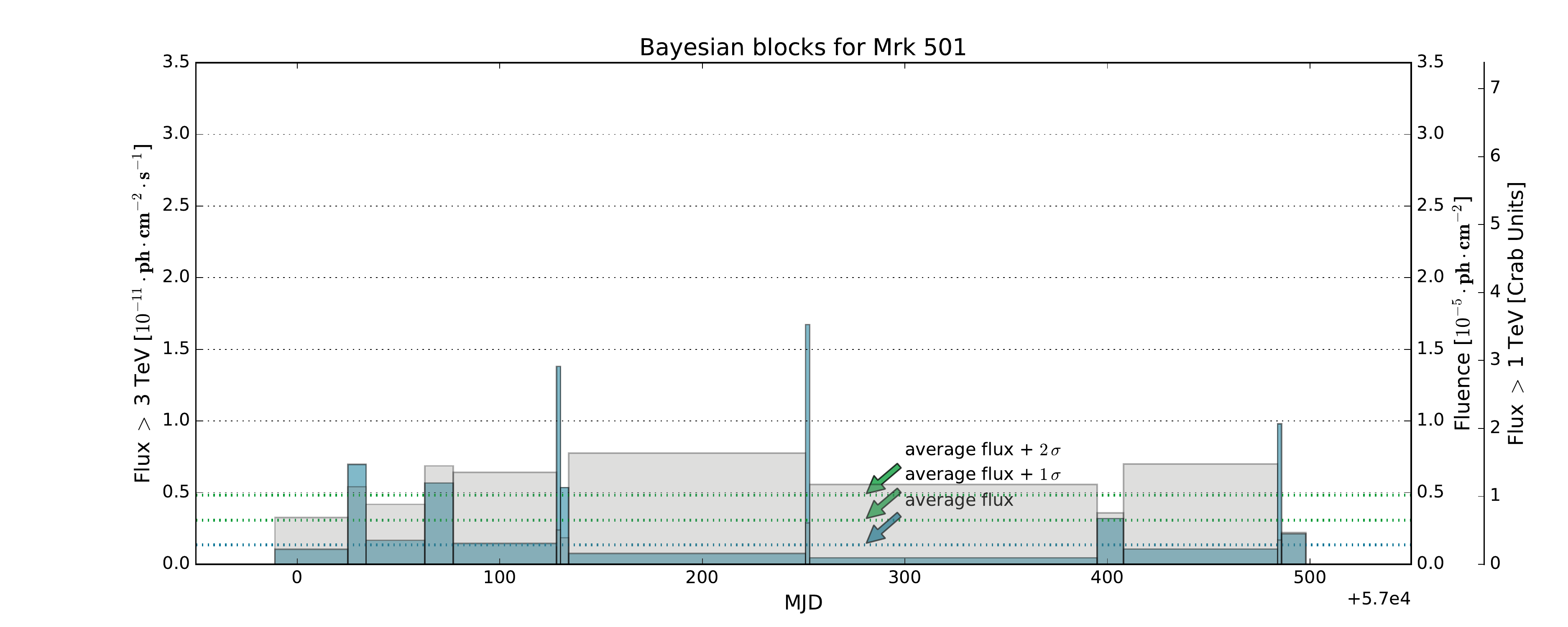}
\caption{Flare states for Mrk 501 vs threshold. The blue dotted line represent the average flux $\sim0.3$ CU (Crab Units); the green dotted lines represent the peak selection thresholds: \textit{average flux}, \textit{average flux +} $\mathit{1\sigma}$, \textit{average flux +} $\mathit{2\sigma}$. The bottommost plot shows the long case, the three upper plots show short case for \textit{average flux}, \textit{average flux +} $\mathit{1\sigma}$, \textit{average flux +} $\mathit{2\sigma}$ respectively. The left axes represent the units of the fluxes, the right-right axis represent the fluxes in corresponding CU. The right-left axes represent the units of fluences shown as shaded grey areas.}
\label{fig:mrk501threshold}
\end{figure}

The signal time PDF shape is extracted directly from the $\gamma$-ray light curve assuming a proportionality between the $\gamma$-ray and the $\nu$ fluxes (see~\Cref{fig:mrk421threshold,fig:mrk501threshold}). 

The test statistics (TS) is evaluated by generating pseudo-experiments simulating background and signal around the considered source:
\begin{equation}
TS = 2(\mathrm{ln}(L^{max}_{s+b})-\mathrm{ln}(L_{b}))
\end{equation}

  
Contrarily to other analyses~\cite{blazarsfive,xrb}, it is assumed that neutrinos are emitted all along the LC and not only in the selected peaks. If the only high peaks selected with neutrinos injected solely there, it gives the flux outside of those peaks being artificially treated as zero which subsequently can raise the loss of neutrinos. Therefore, the derived $N_{S}$ required for discovery for the selected peaks is then rescaled as if like neutrinos injected on all flares.
Conversion from $N_{S}$ to $F$, the equivalent source flux, is done through the acceptance of the detector.

\begin{figure}[!htb]
\centering
\begin{minipage}{0.5\textwidth}
  \centering
\begin{tikzpicture}
    \node[anchor=south west,inner sep=0] (image) at (0,0) {\includegraphics[width=1.06\textwidth] {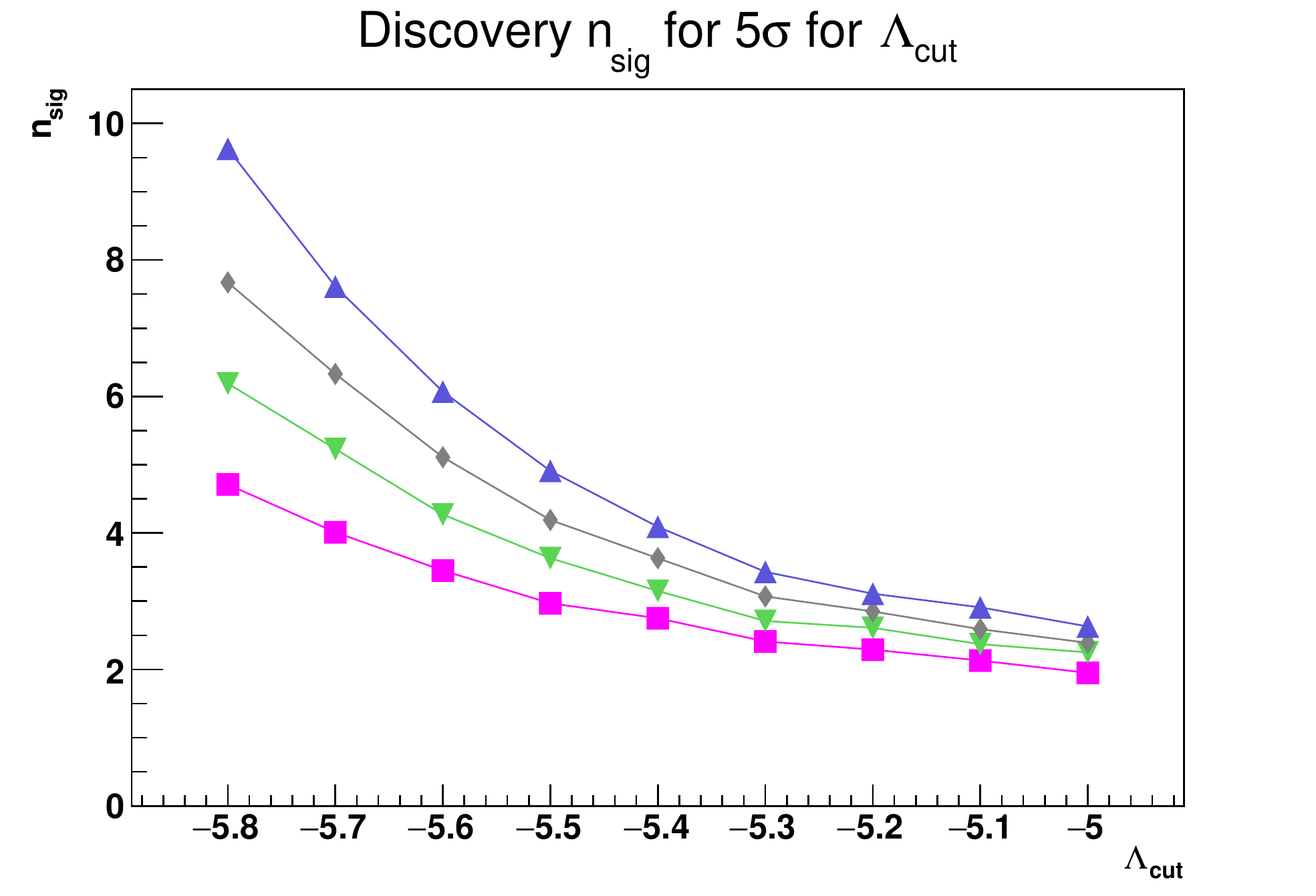}}; 
    \begin{scope}[x={(image.south east)},y={(image.north west)}]
        \node[text width=8cm, minimum height=3cm,minimum width=8cm] at (0.785,0.21) { \textbf{ \textcolor{red}{ 
        \normalsize{PRELIMINARY}  
       } } };
       \end{scope}
   \end{tikzpicture}
\end{minipage}%
\begin{minipage}{0.5\textwidth}
  \centering
  \begin{tikzpicture}
     \node[anchor=south west,inner sep=0] (image) at (0,0) {\hspace*{0.2cm}\includegraphics[width=1.06\textwidth]{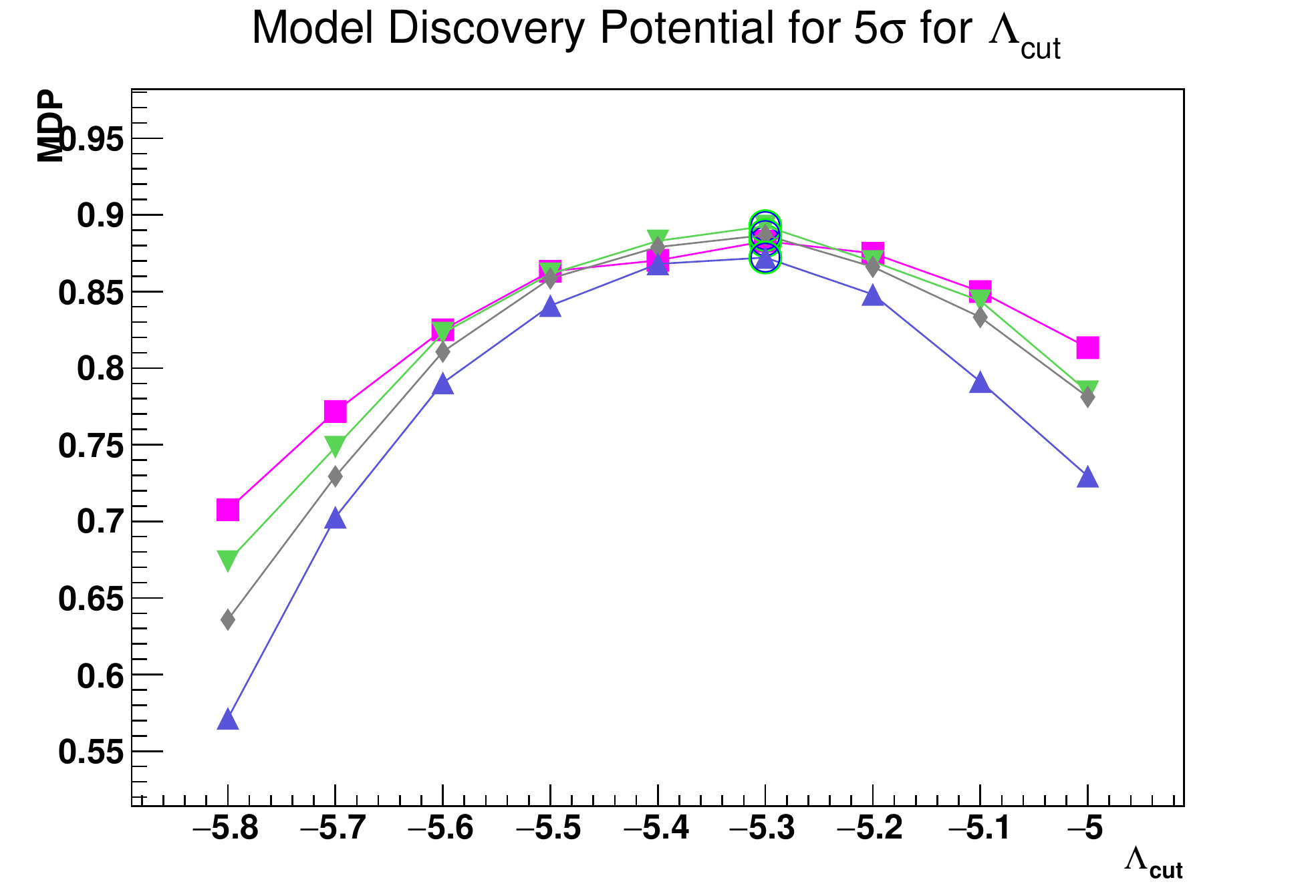}};
    \begin{scope}[x={(image.south east)},y={(image.north west)}]
        \node[text width=8cm, minimum height=3cm,minimum width=8cm] at (0.785,0.21) { \textbf{ \textcolor{red}{
        \normalsize{PRELIMINARY}  
       } } };
       \end{scope}
   \end{tikzpicture}
\end{minipage}
\caption{Examples of discovery power (left) and MDP (right) at $5\sigma$ level vs $\Lambda_{cut}$ for Mrk 501 for $E^{-1.0} \, \mathrm{exp}\left(-E \slash 1 \, \mathrm{PeV}\right)$ (pink), $E^{-2.0}$ (green), $E^{-2.25}$ (grey), $E^{-2.5}$ (blue) spectra in case of all flare states selected. The light green color circles represent the MDP$^{5\sigma}_{max}$ values.}
\label{fig:nsigandmdp}
\end{figure}

Cuts on the cosine of the zenith angle of the reconstructed events $\mathrm{cos(\theta)>-0.1}$ and on the direction of the reconstructed events $\beta<1.0^{\circ}$ are used to improve the signal-to-noise ratio. The $\Lambda_{cut}$ is optimized for each source on the basis of maximizing Model Discovery Potential (MDP) for $5\sigma$ level for each $\nu$ spectrum (see Fig.~\ref{fig:nsigandmdp}). MDP is a probability to make a discovery assuming that the model is correct~\cite{grb2013}.

\clearpage

\clearpage

\section{Results.}

\begin{figure}[!htb]  
  \begin{minipage}{0.5\textwidth}
  \centering
  \begin{tikzpicture}
    \node[anchor=south west,inner sep=0] (image) at (0,0) {\includegraphics[width=1.2\textwidth] {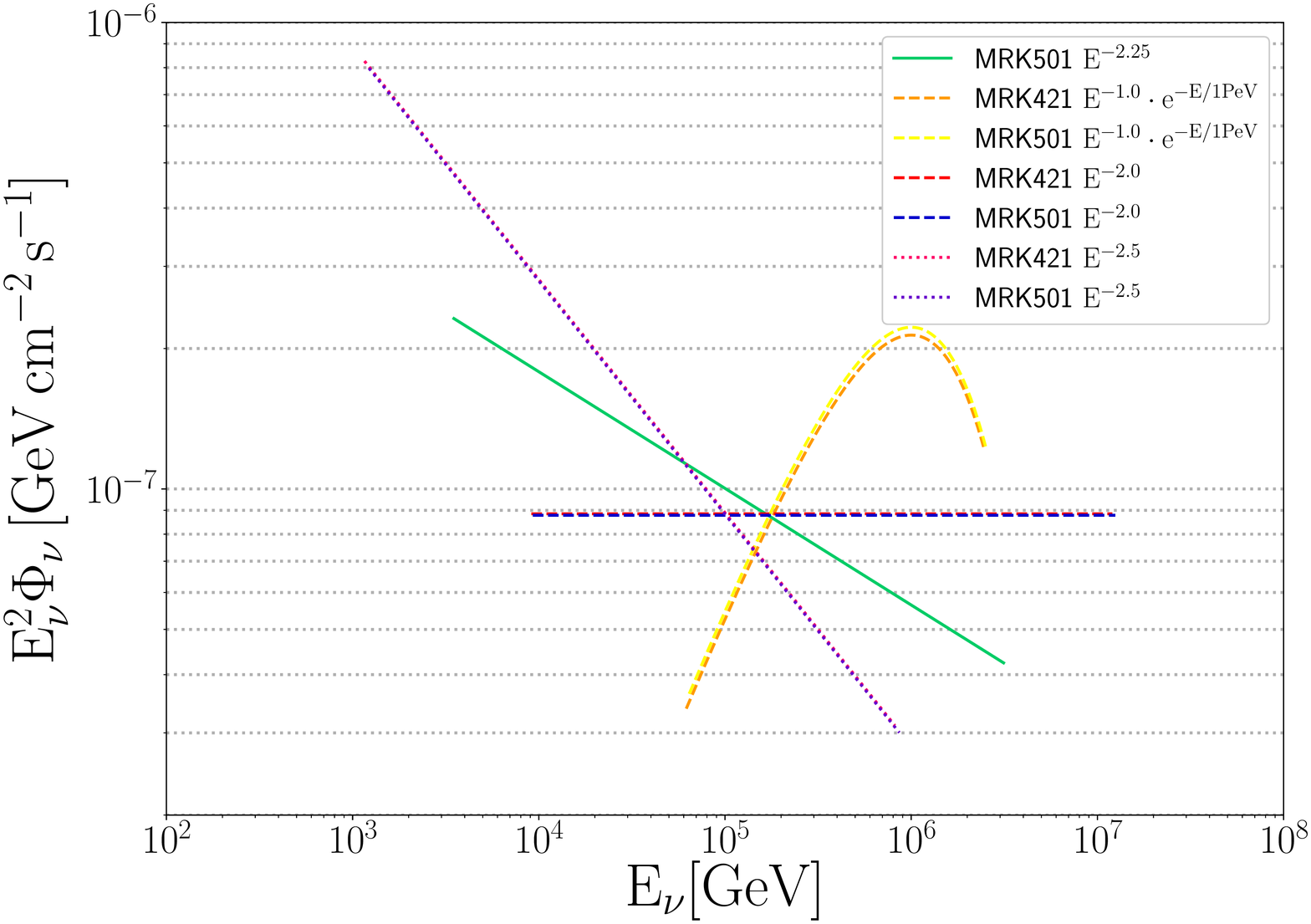} };
    \begin{scope}[x={(image.south east)},y={(image.north west)}]
        \node[text width=8cm, minimum height=3cm,minimum width=8cm] at (0.76,0.1518) { \textbf{ \textcolor{red}{ 
        \normalsize{PRELIMINARY}  
       } } };
       \end{scope}
   \end{tikzpicture} 
  \end{minipage}%
  \begin{minipage}{0.5\textwidth}
  \centering 
  \vspace{-0.98cm}  
  \begin{tikzpicture}
    \node[anchor=south west,inner sep=0] (image) at (0,0) 
{\hspace*{+1.0cm}\includegraphics[width=0.835\textwidth]{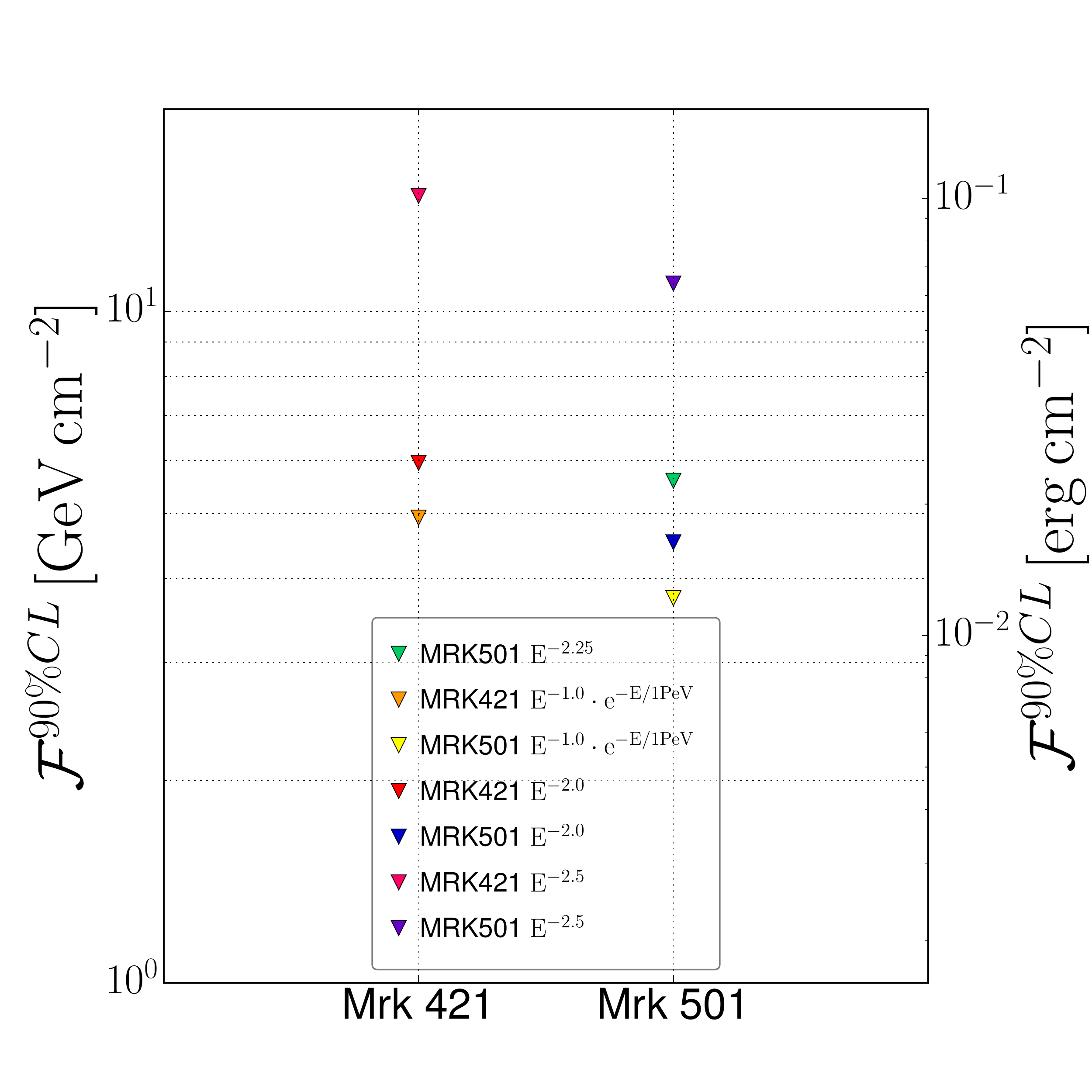} }; 
\begin{scope}[x={(image.south east)},y={(image.north west)}]
        \node[text width=8cm, minimum height=3cm,minimum width=8cm] at (0.88,0.865) { \textbf{ \textcolor{red}{ 
        \normalsize{PRELIMINARY}  
       } } };
       \end{scope}
   \end{tikzpicture} 
  \end{minipage}
\caption{Neutrino energy flux sensitivities (left) and neutrino fluence sensitivities (right) at $90\%$ C.L. obtained in the analysis with the all neutrino energy spectra and for cases of all flare states selected and flaress with \textit{average flux}+$\mathit{2\sigma}$ threshold selected respectively.}
\label{fig:ULfluxANDfluence}
\end{figure}

The sensitivities at $90\%$ C.L. on neutrino energy fluxes and fluences obtained with $90\%$ C.L. sensitivity fluxes for optimum $\Lambda$ cut values for each spectrum and gathered in Fig.~\ref{fig:ULfluxANDfluence}.

\begin{figure}[!htb]  
  \begin{minipage}{0.5\textwidth}
  \centering
  \begin{tikzpicture}
    \node[anchor=south west,inner sep=0] (image) at (0,0) {\includegraphics[width=1.0\textwidth]{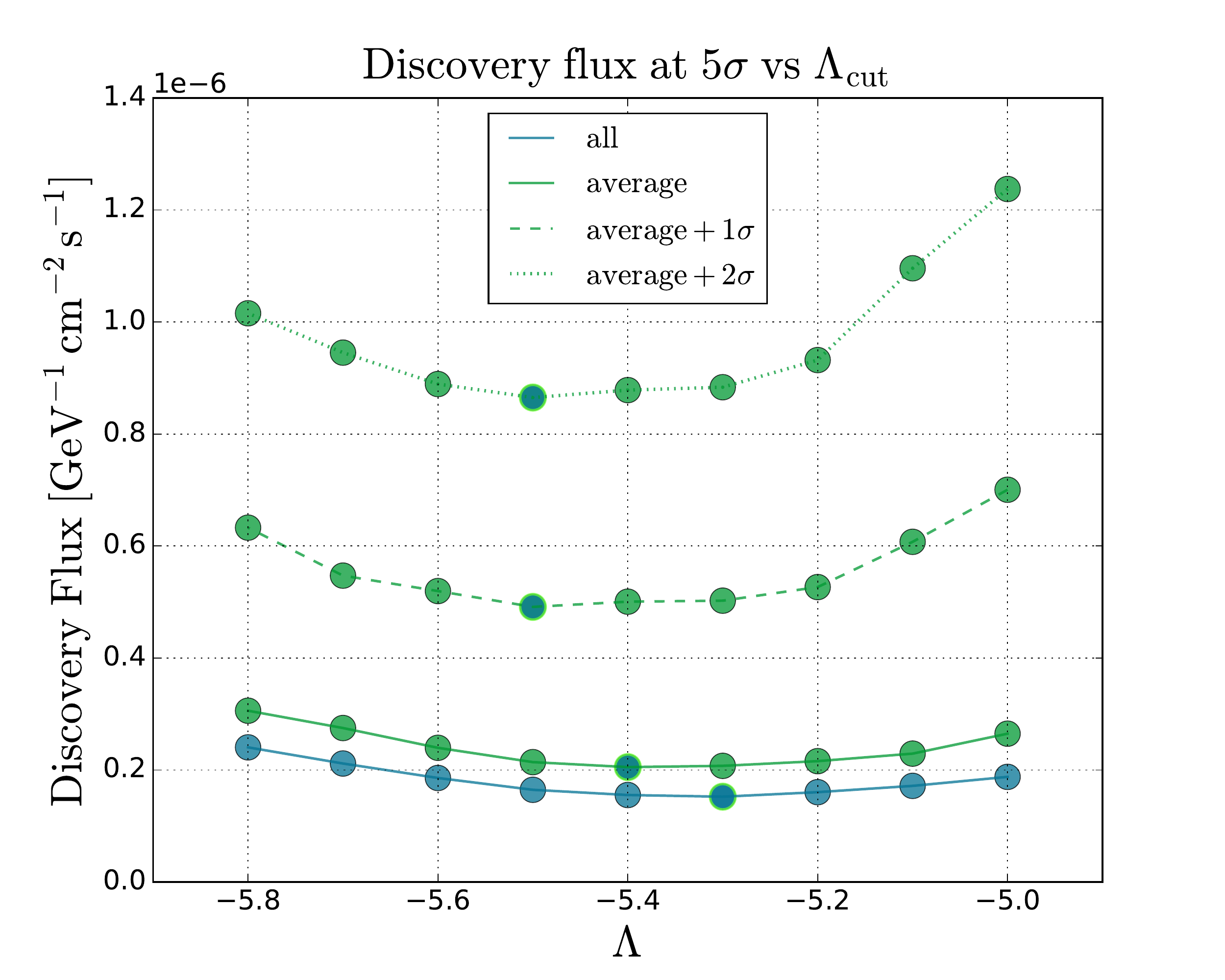} }; 
    \begin{scope}[x={(image.south east)},y={(image.north west)}]
        \node[text width=8cm, minimum height=3cm,minimum width=8cm] at (0.81,0.148) { \textbf{ \textcolor{red}{
        \normalsize{PRELIMINARY}  
       } } };
       \end{scope}
   \end{tikzpicture} 
  \end{minipage}%
  \begin{minipage}{0.5\textwidth}
  \centering
  \begin{tikzpicture}
    \node[anchor=south west,inner sep=0] (image) at (0,0) {\includegraphics[width=1.00\textwidth]{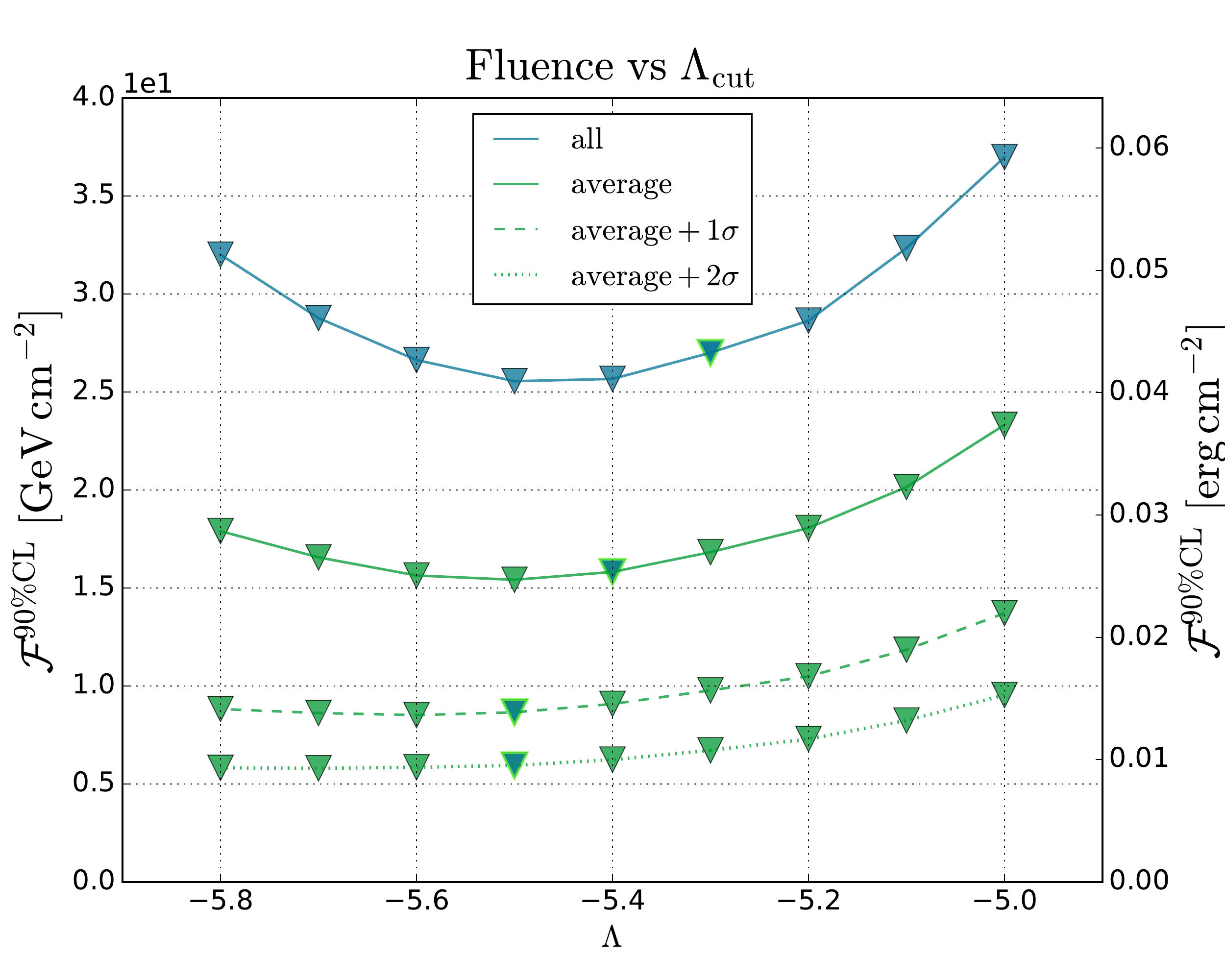} };
    \begin{scope}[x={(image.south east)},y={(image.north west)}]
        \node[text width=8cm, minimum height=3cm,minimum width=8cm] at (0.8,0.15) { \textbf{ \textcolor{red}{
        \normalsize{PRELIMINARY}  
       } } };
       \end{scope}
   \end{tikzpicture}  
\end{minipage}
\caption{Examples of discovery fluxes comparison at $5\sigma$ level (left) and neutrino fluence sensitivities at $90\%$ C.L. (right) vs $\Lambda$ cut for Mrk 421 for $E^{-2.0}$ spectrum with different peaks selection thresholds. Light green color circles represent the values with $\Lambda$ cut that maximizes $\mathrm{MDP}^{5\sigma}$. }
\label{fig:discfluxandfluence}
\end{figure}

The lowest flux required for discovery was obtained with the case of all flare states selected (see Fig.~\ref{fig:discfluxandfluence}). In contrast, the lowest fluence was obtained with \textit{average flux}+$\mathit{2\sigma}$ threshold (see Fig.~\ref{fig:discfluxandfluence}). Usage of this threshold instead of all flare states makes possible to set a better upper limits in the absence of discovery.

\section{Conclusion}
The HAWC detector operates nearly continuously and it is currently the most sensitive wide FOV $\gamma$-ray telescope in the very promising HE band from 100 GeV to 100 TeV. Therefore, it opens prospects to study the most energetic astrophysical phenomena in the Universe as well as to understand the mechanisms that power them and endeavor to break the mystery of their origin. Taking into account the flare timing information given by $\gamma$-ray observations should improve the efficiency of the search for a $\nu$ counterpart with ANTARES. The next generation KM3NeT neutrino telescope~\cite{km3net} will provide more than an order of magnitude improvement in sensitivity~\cite{km3netNEUTRINO2018talk}; therefore, such sources are promising candidates as HE $\nu$ emitters for an improved future time-dependent search.


\end{document}